\documentclass[10pt,pra,twocolumn,superscriptaddress,showpacs,floatfix]{revtex4}

\usepackage{graphicx}
\usepackage{amssymb}
\usepackage{times}
\usepackage{amsmath}
\usepackage{amsthm}

\begin{document}

\title{Continuous-Variable Quantum Key Distribution with Entanglement in the Middle}

\author{Christian Weedbrook}\email{christian.weedbrook@gmail.com} \affiliation{Center for Quantum Information and Quantum Control, Department of Electrical and Computer Engineering and
Department of Physics, University of Toronto, Toronto, M5S 3G4, Canada}

\date{\today}

\begin{abstract}

We analyze the performance of continuous-variable quantum key distribution protocols where the entangled source originates not from one of the trusted parties, Alice or Bob, but from the malicious eavesdropper in the middle. This is in contrast to the typical simulations where Alice creates the entangled source and sends it over an insecure quantum channel to Bob. By using previous techniques and identifying certain error correction protocol equivalences, we show that Alice and Bob do not need to trust their source, and can still generate a positive key rate. Such a situation can occur in a quantum network where the untrusted source originated in between the two users.


\end{abstract}


\maketitle

\section{Introduction}

Quantum key distribution (QKD)~\cite{Sca09} provides a way for two people, Alice and Bob, to communicate secretly over an insecure quantum channel. The continuous-variable version of QKD~\cite{Weedbrook2011} provides an exciting alternative to the original discrete-variable schemes. Continuous variables offer higher detection efficiencies, off-the-shelf lasers for sources, and the potential to be conveniently integrated into current telecommunication systems. Theoretically, the use of Gaussian resources, operations, and measurements, offer a simple way of analyzing the security of such protocols.

Significant progress has been made in continuous-variable QKD over the years~\cite{Weedbrook2011}. More recent results include, the first experimental demonstration of entanglement-based continuous-variable QKD~\cite{Madsen2011}, as well as finite-size key effects with composability using squeezed states~\cite{Furrer2011}, and improved higher efficiency error correcting codes~\cite{Jouguet2011}. In the performance analysis of continuous-variable QKD protocols thus far, it is the typical assumption that the source of the QKD protocol is created by one of the honest parties, e.g., Alice, and independent of the eavesdropper, Eve. However, having Eve being the source of the entanglement has long been a staple of theoretical discrete-variable QKD~\cite{Ekert91}, especially with respect to security proofs~\cite{Lo99,Mayers1998,Acin2007,Lo2011,Sca09}, and practical demonstrations~\cite{Waks2002,Ma2007,Erven2008}. On the other hand, explicitly calculating the secret key rates in such a situation has not been considered in continuous-variable QKD, even though benefits such as higher tolerance to loss have been observed in discrete-variable QKD~\cite{Waks2002,Ma2007}.

In this paper, we ask the question: ``does having the entangled source originating from the middle, improve the performance of continuous-variable QKD protocols as it does in certain discrete-variable protocols?". Hence, we consider the continuous-variable situation where Eve is placed in the middle between Alice and Bob, and is given full control of the creation of the Gaussian entangled resource. Using previous analytical techniques (e.g., see~\cite{GarciaPatron2007}), we show that in such a situation, a secure key can still be generated between Alice and Bob. Furthermore, our analysis uncovers the relationships and equivalences between the error correction protocols, direct~\cite{Grosshans2002} and reverse reconciliation~\cite{Grosshans2003}. For example, direct reconciliation using coherent states and homodyne detection is shown to be equivalent to reverse reconciliation using squeezed states and heterodyne detection when the entanglement is placed in the middle. In such a situation, from one perspective, direct reconciliation is improved and can beat the $3$~dB loss limit, while from another, it is simply a worse off version of reverse reconciliation (e.g., reverse reconciliation with excess channel noise). This highlights the differences between discrete and continuous-variable QKD, particularly in how loss on the channel affects each of them.

This paper is structured as follows. In Sec.~II, we describe the entanglement in the middle scheme followed by the equivalences of the error correcting protocols in Sec.~III. In Sec.~IV, we derive the secret key rates of various protocols where the entanglement originated from Eve in the middle. We follow this with a discussion of our results in Sec.~V and offer concluding remarks in Sec.~VI.

%

%

%
\begin{figure}[!ht]
\begin{center}
\includegraphics[width=8cm]{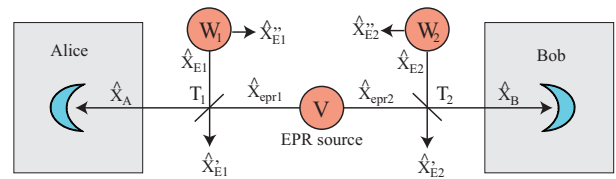}
\caption{Schematic of a continuous-variable QKD protocol where the entangled source of variance $V$ (potentially created by Eve) is placed in the middle between Alice and Bob. Eve's attack consists of two entangling cloner attacks on either side of the source (see text for further details). Alice and Bob can either perform homodyne or heterodyne detection, using either direct or reverse reconciliation. Note that the usual situation of a trusted source (i.e., a source originating from Alice's station) can simply be recovered by setting $T_1=1$.}\label{schematic}
\end{center}
\end{figure}

\section{Entanglement in the Middle}

We begin by first describing the entanglement in the middle scheme. An entangled Gaussian resource is placed in the middle between Alice and Bob, see Fig.~\ref{schematic}. This resource is to be used by Alice and Bob to generate a secure key which can then be used to encrypt data~\cite{Sca09}. In principle, this resource could have been created by some third party, say Charlie, in installing an entangled source between two users in a quantum network. However, it is better to err on the side of caution and assume that Eve could have also created this entangled state. We also assume that this entangled source is Gaussian. This is because the source is usually a Gaussian state and can thus maximize the Shannon mutual information and from an eavesdropping point of view, a Gaussian attack maximizes the eavesdropper's extractable information~\cite{Navascues2005,Gar06}. In her attack, Eve perfectly replaces the quantum channel between Alice and Bob with her own quantum channel where the loss is simulated by two separate beam splitters with transmissions $T_1$ and $T_2$, where $T_i \in [0, 1]$. When these two transmissions are symmetric, i.e., $T_1=T_2$, we say that the entanglement is in the middle of Alice and Bob.

Eve's EPR state has two entangled modes described in the Heisenberg picture as $\hat{X}_{epr1}$ (sent to Alice) and $\hat{X}_{epr2}$ (sent to Bob). Such a state is created by combining two orthogonal squeezed states, $\hat{X}_{s1}$ and $\hat{X}_{s2}$, on a $50/50$ beam splitter.  i.e.,
\begin{align}
\hat{X}_{epr1} &= (\hat{X}_{s1}+\hat{X}_{s2})/\sqrt{2},\\\nonumber
\hat{X}_{epr2} &= (\hat{X}_{s1}-\hat{X}_{s2})/\sqrt{2}.
\end{align}
Also $V$ is the symmetrized variance of each of the two entangled modes, i.e., $V:= V(\hat{X}_{epr1}) = V(\hat{X}_{epr2})$. We also assume Eve's attack is perfect, i.e., no unknown noise on her input and output states. Note that when $T_1=1$ we can recover the usual QKD situation where Alice creates the entangled state safely at her station. For each of these beam splitters, Eve performs a collective Gaussian attack~\cite{Navascues2005,Gar06,Pir08}. We assume Eve uses such an attack in our new protocol, because it was shown, up to a suitable symmetrization of the protocols, that this type of attack is the best attack allowed by the laws of quantum physics for standard direct and reverse reconciliation protocols~\cite{Ren09}. It consists in Eve interacting her independent ancilla modes with Alice and Bob's resultant modes for each run of the protocol in such a way to generate a memoryless Gaussian channel.

The most common example of a collective Gaussian attack is the entangling cloner~\cite{Gross03}. This consists in Eve preparing (for each of the two beam splitter attacks) ancilla modes $\hat{X}_{E}$ and $\hat{X}''_{E}$ from an entangled Gaussian state with variance $W$. Eve keeps one mode $\hat{X}''_E$ and injects the other mode $\hat{X}_{E}$ into the unused port of the beam splitter, leading to the output mode $\hat{X}'_{E}$. These operations are repeated identically and independently for each of the signal modes sent out to Alice and Bob. Eve's output modes are then stored in a quantum computer and detected collectively at the end of the protocol. Eve's final measurement is optimized based on Alice and Bob's classical communications.

Finally, we note that in continuous-variable QKD, in order for Alice and Bob to perform their quadrature measurements via homodyne or heterodyne detection, they need a reference pulse known as a local oscillator. This allows them to correctly measure a quadrature with respect to the other orthogonal quadrature. However, the local oscillator can be simply thought of as a classical signal and can therefore be considered an authenticated classical signal~\cite{Haseler2008}. So in principal, when the source originates from the middle, Alice can first create a local oscillator and send it onto Bob.

\begin{figure}[!ht]
\begin{center}
\includegraphics[width=8.5cm]{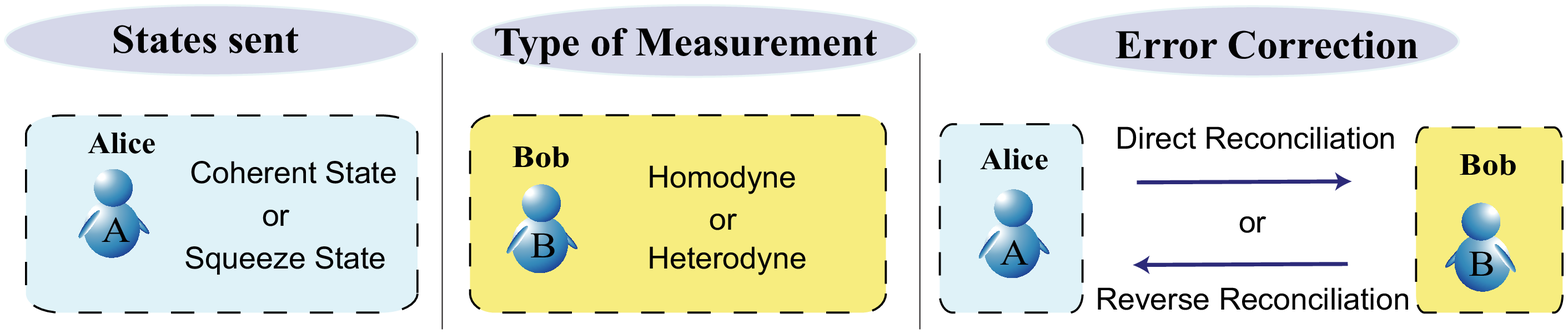}
\caption{An example of a family of eight protocols~\cite{Gar09} in continuous-variable QKD. Alice can choose between two states to send to Bob, who can in turn choose between two types of measurements. Finally there are two one-way error correction protocols Alice and Bob can use.}\label{equiv4}
\end{center}
\end{figure}

\section{Equivalences Between Direct and Reverse Reconciliation}

The first result of this paper is to show that a number of equivalences occur between the error correction protocols direct reconciliation~\cite{Grosshans2002} and reverse reconciliation~\cite{Grosshans2003}, when entanglement is placed in the middle. To understand this, it is first helpful to understand how the entanglement-based picture works in continuous variables~\cite{GarciaPatron2007}. Suppose Alice and Bob share an entangled bi-partite Gaussian state. Then if Alice performs homodyne detection (and therefore gets one measurement result, $q$ or $p$) on her half of the EPR pair then Bob's half collapses to a squeezed state. On the other hand, if Alice performs heterodyne detection (and thus gets two measurement results, $q$ and $p$), Bob's mode collapses to a coherent state. So in the prepare-and-measure scheme, it is as though Alice had initially sent Bob a randomly displaced squeezed (coherent) state, respectively. As for the reconciliation protocols, when we talk about direct (reverse) reconciliation it means that Bob (Alice) is trying to estimate what Alice (Bob) has measured. Hence, in direct (reverse) reconciliation Alice (Bob) is the reference point. In total then, with the state preparation, detection and error correction, we end up with a family of eight protocols~\cite{Gar09}, i.e., one could choose either squeezed or coherent states with either direct or reverse reconciliation using either homodyne or heterodyne detection (cf. Fig.~\ref{equiv4}).

We naturally find certain equivalences between the eight protocols, when entanglement is in the middle, based on the type of measurement Alice and Bob perform. To work out the equivalences simply determine the number of measurement outcomes Alice and Bob get for each run of the protocol, e.g., if both perform homodyne each will have one measurement outcome, whereas with heterodyne detection both will have two. If they each have the same number of measurement outcomes, then direct and reverse reconciliation protocols will be the same. For example, if the total number of measurements for each are two, meaning Alice prepared coherent states and Bob performed heterodyne detection, then it really doesn't matter who we label as Alice and who we label as Bob. The protocol becomes symmetric as they both have two measurement outcomes, and hence who is the reference point is arbitrary and so, direct reconciliation equals reverse reconciliation.

The explicit equivalences between the protocols, when entanglement is placed in the middle, are as follows:
\begin{itemize}
  \item Coherent states and homodyne with direct (reverse) is the same as squeezed states and heterodyne with reverse (direct).
  \item Direct and reverse reconciliation are equivalent for squeezed states and homodyne and also for coherent states and heterodyne.
\end{itemize}
In this paper, we show that out of the family of eight protocols in total, coherent states and homodyne with direct, or equivalently, squeezed state and heterodyne with reverse, outperform all other protocols when the source is placed in the middle. We will now calculate the secret key rates for the various protocols using previously established techniques, e.g., see \cite{GarciaPatron2007}.

\section{Secret Key Rates}

We will limit our calculations to direct reconciliation, as the calculations for reverse reconciliation can be automatically derived through the previously mentioned equivalences. The secret key rate for direct reconciliation is defined as
\begin{align}
K = S(A:B) - S(A:E)\label{eq: secret key rate},
\end{align}
where $S(A:B)$ and $S(A:E) = S(E) - S(E|A)$ are the mutual informations between Alice and Bob and Alice and Eve, respectively. We now go about calculating these quantities for the cases of Alice creating squeezed states and coherent states with Bob using either homodyne or heterodyne detection~\cite{C.Weedbrook2004}. For more details of these calculations involving correlation matrices and symplectic spectra see~\cite{Weedbrook2011,GarciaPatron2007}. Note also that we assume Alice and Bob have perfect ($100$~\% efficient) homodyne and heterodyne detectors. First we write Alice and Bob's covariance matrix~\cite{Weedbrook2011}, which is given by
\begin{align}\label{eq: covariance matrix}
\gamma_{AB} =
\left(
  \begin{array}{cc}
    a \textbf{I} & c \textbf{Z} \\
    c \textbf{Z} & b \textbf{I} \\
  \end{array}
\right),
\end{align}
where $\textbf{I}$ and $\textbf{Z}$ are the usual Pauli matrices and $a = T_1 V + (1-T_1) W_1$, $b = T_2 V + (1-T_2) W_2$, and $c = \sqrt{T_1} \sqrt{T_2} \sqrt{V^2-1}$. This seemingly  innocuous matrix will form the basis for most of our calculations. In fact, it is this matrix which takes into account Eve being in the middle (along with the requirements of the effective channel, $T=T_1 T_2$, which will be used when plotting), because typically $a = V$. 

Note that the value $T$ can be calculated by allowing an arbitrary mode $\hat{X}_1$ to go through two different beam splitters one after the other. For example, after the first beam splitter we have $\hat{X}_2 = \sqrt{T_1} \hat{X}_1 + \sqrt{1-T_1}\hat{N}_1$ and after the second beam splitter $\hat{X}_3 = \sqrt{T_2} \hat{X}_2 + \sqrt{1-T_2}\hat{N}_2$. Back substituting and calculating the variances, gives the final required relation, $T=T_1 T_2$, where $\hat{N}_i$ are the noise terms from the unused port of the beam splitters.

\subsection{Squeezed states}

Alice and Bob's mutual information, for squeezed states~\cite{Cerf2001,Usenko2011} with homodyne detection, is given by $\frac{1}{2} \log_2 (V_A/V_{A|B})$, i.e.,
\begin{align}
S(A:B) &= \frac{1}{2} \log_2 \Big(\frac{a}{a - c^2/b} \Big),
\end{align}
for heterodyne is given by
\begin{align}
S(A:B) &= \frac{1}{2} \log_2 \Big(\frac{a}{a - c^2/(b+1)} \Big).
\end{align}
We will now calculate Eve's mutual information with Bob, $S(A:E) = S(E) - S(E|A)$, which is exactly the same for both homodyne and heterodyne detection. This is because, as we are using direct reconciliation, it is conditioned on Alice's measurement results and not Bob's. This simplification illustrates the beauty of the entanglement-based scheme. Note that because Eve provides a purification of Alice and Bob's density matrix we can write $S(E) = S(AB)$. Therefore, we have $S(AB) = G[(\lambda_1-1)/2]+ G[(\lambda_2-1)/2]$ where
\begin{align}
G(x) = (x+1)\log_2 (x+1) - x\log_2 x,
\end{align}
and where the symplectic eigenvalues are given by
\begin{align}
\lambda^2_{1,2} = \frac{1}{2} [\Delta \pm \sqrt{\Delta^2 - 4 D^2}],
\end{align}
where $\Delta = a^2 + b^2 + c^2$ and $D = ab-c^2$~\cite{Weedbrook2011,GarciaPatron2007}. Again using the same purification argument we can write $S(E|A) = S(B|A)$. Again to calculate $S(B|A)$ we calculate Bob's correlation matrix conditioned on Alice's measurement outcome $x_a$ which is calculated using~\cite{Weedbrook2011}
\begin{align}
\gamma^{x_a}_B = \gamma_B - \sigma^T_{AB} (X \gamma_A X)^{-1} \sigma_{AB},
\end{align}
where the inverse is a pseudoinverse and the matrix $X = [1, 0; 0, 0]$. We find that $S(B|A) = G[(\lambda_3-1)/2]$ where $\lambda^2_3 = b (b-c^2/a)$. We now have enough information to numerically calculate and plot the secret key rate given in Eq.~(\ref{eq: secret key rate}), but first we calculate the coherent state case.

\subsection{Coherent states}

Alice and Bob's mutual information, for coherent states with homodyne detection, is given by
\begin{align}
S(A:B) = \frac{1}{2} \log_2 \Big(\frac{a+1}{a+1 - c^2/b} \Big),
\end{align}
and for heterodyne detection it is
\begin{align}
S(A:B) &=  \log_2 \Big(\frac{b+1}{b+1 - c^2/(a+1)} \Big).
\end{align}
As before, the next step is to calculate Eve's mutual information with Alice. Now in the previous case of squeezed states, we only needed to calculate one expression for Eve's mutual information with Alice which could then be used for both the homodyne and heterodyne detection cases. However, this is not possible in the case of coherent states, as Alice's measurement is different depending on whether Bob does homodyne or heterodyne detection. This is because Alice discards one of her measurement results during the sifting phase of homodyne and keeps both during heterodyne. Therefore two separate calculations needs to be performed. Firstly, the homodyne detection case. To create a coherent state, Alice performs heterodyne detection on her mode using a $50/50$ beam splitter (BS) which introduces vacuum noise denoted by system $C$. This interaction on the initial correlation matrix $\gamma_{A_0C_0B}$ can be described by the following symplectic transformation~\cite{Weedbrook2011} $\gamma_{ACB} = [S_{AC}^{BS} \otimes I_B]^T \gamma_{A_0C_0B} [S_{AC}^{BS} \otimes I_B]$. Using the previous purification arguments, we find that $S(E|A) = S(BC|A)$ as, after Alice's measurement, the system $BCE$ is pure. The correlation matrix of the system $BC$ conditioned on Alice's measurement result is calculated to be
\begin{figure}[!ht]
\begin{center}
\includegraphics[width=8cm]{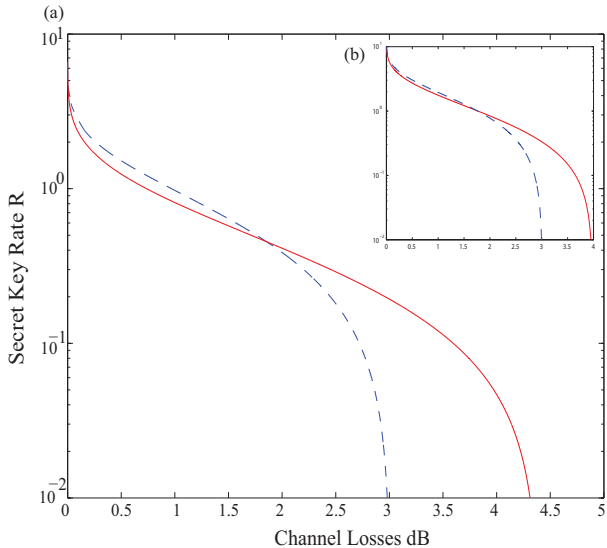}
\caption{Plot of direct reconciliation using homodyne detection with (a) coherent states and (b) squeezed states. Solid (red) line: entanglement in the middle; Dashed (blue) line: entanglement at Alice's station. By giving Eve control of the entangled source and by placing it in the middle of Alice and Bob we are still able to generate a secure key. Due to the equivalences in the error correction protocols, this entanglement in the middle scenario can be seen as a direct reconciliation protocol beating the 3~dB loss limit using coherent states or alternatively, a reverse reconciliation protocol with excess channel noise.}\label{2_way_comp2}
\end{center}
\end{figure}
\begin{align}\nonumber
\gamma_{BC}^{x_a} =
\left(
  \begin{array}{cccc}
    b-c^2/(a+1) & 0 & \sqrt2 c/(a+1) & 0 \\
    0 & b & 0 & -c/\sqrt2 \\
    \sqrt2c/(a+1) & 0 & 2a/(a+1) & 0 \\
    0 & -c/\sqrt2 & 0 & (a+1)/2 \\
  \end{array}
\right).
\end{align}
The conditional von Neumann entropy can be written as $S(BC|A) = G[(\lambda_3-1)/2]+G[(\lambda_4-1)/2]$ where
\begin{align}
\lambda^2_{3,4} = \frac{1}{2} [A \pm \sqrt{A^2 - 4B}],
\end{align}
with $A = (a+bD+\Delta)/(a+1)$ and $B = D(b+D)/(a+1)$ with $D$ and $\Delta$ defined previously.

We now calculate Eve's mutual information in the case of Bob performing heterodyne detection. Using $\gamma_B^{x_a,p_a} = \gamma_B - C(X \gamma_{AC} X)^{-1} C^T$ gives $\gamma_{BC}^{x_a} = [b-c^2/(a+1)] \bf{I}$. The conditional von Neumann entropy is given by $S(E|A) = S(B|A) = G[(\lambda_3-1)/2]$ where $\lambda_3 = b-c^2/(a+1)$. We can now calculate the final secret key rate for the coherent state protocols and plot them with the squeezed states protocols.

\section{Results and Discussion}

Using the previous section's results we can plot the secret key rates for the various protocols over a lossy channel. We find that out of the family of eight protocols in total, coherent states and homodyne with direct, or equivalently, squeezed state and heterodyne with reverse, outperform all other protocols when the source is placed in the middle.
In Fig.~\ref{2_way_comp2}, we plot the cases of homodyne detection for (a) coherent states and (b) squeezed states where the dashed lines indicate the usual case where Alice controls the entangled source, while the solid lines indicate the entanglement in the middle situation. We can see that in both the coherent state and squeezed state cases, even though Eve controls the source, both protocols are still secure. We also notice that the entanglement in the middle schemes can go past the $3$~dB loss limit of direct reconciliation. In the equivalences between direct and reverse reconciliation protocols, this can be looked at as a (coherent state) direct reconciliation protocol beating the $3$~dB loss limit or, alternatively, a (squeeze state) reverse reconciliation protocol performing poorly, i.e., with (equivalent) additional excess channel noise on the channel. Remember that the squeezed state protocol using reverse reconciliation does not have a loss limit and, for a lossy channel, is secure for all values of channel transmissions. We note that the heterodyne cases are not plotted here but are also secure when the source originates from Eve. Also the effect of excess noise on the quantum channel, has the typical effect of reducing the security of the protocol as a function of the excess noise value.

The analysis in this paper highlights the differences between discrete-variable QKD and continuous-variable QKD. This is because in discrete variables, placing the entanglement in the middle allows the QKD protocol to tolerate higher levels of loss (in fact, the distance can be double). However, in continuous variables, we do see an advantage in tolerance to loss but because of the symmetry in error correction protocols, one could look at it as another protocol performing badly. A kind of the glass is half full half empty scenario. Perhaps the main difference lies in the definition of loss. In discrete variables, loss simply means the photon did not arrive and was lost to the environment. Whereas, in continuous variables loss means additional Gaussian noise on the quantum channel, and are thus unable to simply postselect out the times when loss occurs.

\section{Conclusion}

In conclusion, we have analyzed the performance of eight continuous-variable QKD protocols where the source of the protocol originates from the malicious eavesdropper who is in the middle between Alice and Bob. This is in direct contrast to the typical scenario where Alice usually creates the source at her secure station. We showed that a secure key can still be achieved even when Eve controls such a source. We also pointed out the various error correction protocols equivalences that occur when the entanglement is placed in the middle. In terms of possible future research, it would be interesting to look at the entanglement in the middle scheme from the point of view of the postselection error correction protocol~\cite{Sil02}.

\acknowledgments

C.~W. would like to thank Asma Al-Qasimi, Eric Chitambar, Wolfram Helwig, Hoi-Kwong Lo, Norbert L\"utkenhaus, Stefano Pirandola, Tim Ralph for discussions and proof reading of paper. C.~W. acknowledges support from the Ontario postdoctoral fellowship program,
CQIQC postdoctoral fellowship program, CIFAR, Canada Research Chair program,
NSERC, and QuantumWorks.\\

\end{document}